\documentclass{aastex631}%[linenumbers]

\begin{document}
\title{Negative-energy waves in the vertical threads of a solar prominence}
%\correspondingauthor{Jincheng Wang, Dong Li}
%\email{wangjincheng@ynao.ac.cn, lidong@pmo.ac.cn}

\author{Jincheng Wang}
\affiliation{Yunnan Observatories, Chinese Academy of Sciences, Kunming Yunnan 650216, People's Republic of China}
\affiliation{Yunnan Key Laboratory of Solar Physics and Space Science, Kunming 650216, People's Republic of China}

%\author{Xiaoli Yan}
%\affiliation{Yunnan Observatories, Chinese Academy of Sciences, Kunming Yunnan 650216, People's Republic of China}
%\affiliation{Yunnan Key Laboratory of Solar Physics and Space Science, Kunming 650216, People's Republic of China}

\author{Dong Li}
\affiliation{Key Laboratory of Dark Matter and Space Astronomy, Purple Mountain Observatory, Chinese Academy of Sciences, Nanjing 210023, People's Republic of China}
\affiliation{Yunnan Key Laboratory of Solar Physics and Space Science, Kunming 650216, People's Republic of China}

\author{Chuan Li}
\affiliation{School of Astronomy and Space Science, Nanjing University, Nanjing 210023, People's Republic of China}
\affiliation{Key Laboratory of Modern Astronomy and Astrophysics (Nanjing University), Ministry of Education, Nanjing 210023, People's Republic of China}
\affiliation{Institute of Science and Technology for Deep Space Exploration, Suzhou Campus, Nanjing University, Suzhou 215163, People's Republic of China}
\author{Yijun Hou}
\affiliation{National Astronomical Observatories, Chinese Academy of Sciences, Beijing 100101, People's Republic of China}
\affiliation{Yunnan Key Laboratory of Solar Physics and Space Science, Kunming 650216, People's Republic of China}

\author{Zhike Xue}
\author{Zhe Xu}
\author{Liheng Yang}
\affiliation{Yunnan Observatories, Chinese Academy of Sciences, Kunming Yunnan 650216, People's Republic of China}
\affiliation{Yunnan Key Laboratory of Solar Physics and Space Science, Kunming 650216, People's Republic of China}
\author{Qiaoling Li}
\affiliation{Department of Physics, Yunnan University, Kunming 650091, People's Republic of China}
 
\begin{abstract}
Solar prominences, intricate structures on the Sun's limb, have been a subject of fascination due to their thread-like features and dynamic behaviors. Utilizing data from the New Vacuum Solar Telescope (NVST), Chinese H$\alpha$ Solar Explorer (CHASE), and Solar Dynamics Observatory (SDO), this study investigates the transverse swaying motions observed in the vertical threads of a solar prominence during its eruption onset on May 11, 2023. The transverse swaying motions were observed to propagate upward, accompanied by upflowing materials at an inclination of 31 degrees relative to the plane of the sky. These motions displayed small-amplitude oscillations with corrected velocities of around 3-4 km s$^{-1}$ and periods of 13-17 minutes. Over time, the oscillations of swaying motion exhibited an increasing pattern in displacement amplitudes, oscillatory periods, and projected velocity amplitudes. Their phase velocities are estimated to be about 26-34 km s$^{-1}$. An important finding is that these oscillations' phase velocities are comparable to the upward flow velocities, measured to be around 30-34 km s$^{-1}$. We propose that this phenomenon is associated with negative-energy wave instabilities, which require comparable velocities of the waves and flows, as indicated by our findings.
This phenomenon may contribute to the instability and observed disruption of the prominence. By using prominence seismology, the Alfv$\acute{\rm e}$n speed and magnetic field strength of the vertical threads have been estimated to be approximately 21.5 km s$^{-1}$ and 2 Gauss, respectively. This study reveals the dynamics and magnetic properties of solar prominences, contributing to our understanding of their behavior in the solar atmosphere.
\end{abstract}

\keywords{Solar prominences (1519); Solar oscillations (1515); Solar coronal waves (1965)}

\section{Introduction} \label{sec:intro}
Solar prominences or filaments, located in the solar limb or disk, are among the most fascinating structures in the Sun. 
They consist of cool and dense materials that are suspended in the hot and tenuous solar corona, which often lie above the magnetic polarity inversion lines (PILs) \citep{Martin1998SoPh}. Based on their location, they can be classified as active region (AR), intermediate, and quiescent prominences. Generally, intermediate and quiescent prominences, characterized by relatively weak magnetic field strength (a few to tens of Gauss) and greater height (up to 100 Mm), can persist for many weeks at high solar latitudes. In contrast, AR prominences tend to be more dynamic and short-lived, featuring strong magnetic field strengths (tens to hundreds of Gauss) and low height ($<$10 Mm) \citep{Mackay2010SSRv,Vial2015ASSL}.
High-resolution observations have revealed that these structures are resolved into numerous thread-like structures. (e.g., \citealp{Lin2005SoPh,Yan2015RAA}). Solar prominences exhibit various intriguing structures and dynamics, including horizontal and  quasi-vertical threads/motions \citep[e.g.,][]{Chae2008ApJ, Shen2015ApJ}, bubbles \citep{DeToma2008ApJ, Dudik2012ApJ, Chen2021ApJ, Guo2021ApJ}, upflow plumes \citep[e.g.,][]{Berger2008ApJ, Berger2011Natur, Xue2021RAA, Wang2022A&A}, and oscillating motions \citep[e.g.,][]{Oliver2002SoPh, Terradas2002A&A, Ning2009ApJ, Ning2009A&A, Arregui2012LRSP}.
One of controversial questions is about the nature of the quasi-vertical threads in the limb quiescent prominences \citep{Berger2008ApJ,Gibson2018LRSP}. Many authors have proposed that these structures are a pile up of dips in more or less horizontal magnetic field lines, rather than representing true vertical magnetic structures \citep{Chae2008ApJ, Chae2010ApJ, Dudik2012ApJ}. With the determination of velocity vectors in a quiescent prominence, \cite{Schmieder2010A&A} suggested that the vertical structures in limb prominences may not be true vertical magnetic structures in the sky plane. Alternatively, \cite{vanBallegooijen2010ApJ} proposed that many tangled magnetic fields within flux tubes present the magnetic structures of the vertical threads. Through full three-dimensional magnetohydrodynamic simulation, \cite{Jenkins2022NatAs} proposed that vertical structures in the prominences are caused by the magnetic Rayleigh–Taylor instability process. Therefore, the magnetic structure of these vertical threads in the limb prominences remains controversial.

Transverse magnetohydrodynamic (MHD) waves are frequently observed in solar prominences/filaments, typically manifesting as oscillatory motions \citep{Arregui2012LRSP}. Based on their velocity amplitudes, these oscillations can be categorized into large amplitudes (a few tens of km s$^{-1}$) and small amplitudes (a few km s$^{-1}$) \citep{Oliver2002SoPh, Lin2011SSRv, Schmieder2013ApJ, Arregui2012LRSP}. Large amplitude oscillations are often associated with significant solar activities (such as flare, large-scale chromospheric or coronal waves), while small amplitude oscillations are not related to flare activity and their periods, amplitudes, velocities and other characteristics are intimately connected with the local magnetic and thermodynamic structure of prominences \citep{Lin2011SSRv, Arregui2012LRSP}. Previous investigations have shown that small amplitude oscillations exhibit periods ranging from several minutes to tens of minutes, with displacement amplitudes spanning from several hundred kilometers to several thousand kilometers \citep{Okamoto2007Sci, Lin2009ApJ, Li2022SCPMA}.
Using Ca II H line observations from the Hinode/SOT, \cite{Okamoto2007Sci} identified the transverse MHD wave in a solar AR prominence through the vertical oscillating motion of prominence horizontal threads and suggested that these wave could carry sufficient energy to heat corresponding coronal loops. Subsequently, they proposed that the combination of resonant absorption with the Kelvin-Helmholtz instability may be responsible for the dissipation of these wave \citep{Okamoto2015ApJ,Antolin2015ApJ}. 
Recently, fast kink magnetohydrodynamic waves have been detected persistent in a quiescent prominence with no significant damping \citep{Li2022SCPMA}.
The energy carried by the transverse MHD waves may be potential source for coronal heating, which is sufficient for the quiet Sun but not for the AR \citep{Arregui2015RSPTA, VanDoorsselaere2020SSRv}.
However, \cite{Li2022SCPMA} even found that the estimated energy carried by these waves in the prominence was not enough for heating the corona on the quiet Sun.
Therefore, whether these transverse MHD waves in the prominences could heat sufficiently the corona is still under debate. 

With numerical simulations, numerous detailed behaviors of transverse kink waves have been elucidated under various conditions \citep[e.g.,][]{Ruderman2014SoPh,Goossens2014ApJ,Antolin2014ApJ_787L_22,Antolin2017ApJ_836_219,Nakariakov2021SSRv}. When the shear velocity is smaller than the threshold value for the onset of the Kelvin-Helmholtz instability, two kink waves can propagate along the discontinuity. One of the two kink waves becomes a negative-energy wave when the flow velocity exceeds a critical value \citep{Ryutova1988JETP,Ruderman1995JPlPh}. On the other hand, based on MHD wave theories, it is possible to diagnose certain physical parameters (e.g., magnetic field) within prominences/filaments, a method also known as prominence seismology \citep[e.g.,][]{Banerjee2007SoPh,terradas2008ApJ,Soler2015A&A,Vial2015ASSL,Pant2015RAA}. With assumption of a typical density, \cite{Lin2009ApJ} obtained that the Alfv$\acute{\rm e}$n speed and magnetic field strength of threads of a quiescent filament are around 24 km s$^{-1}$ and 2.5 Gauss, respectively. \cite{Li2022SCPMA} derived that the Alfv$\acute{\rm e}$n speed and magnetic field strength of a quiescent prominence threads are around 115 km s$^{-1}$ and 5 Gauss, respectively. \cite{terradas2008ApJ} diagnosed that the lower bound of Alfv$\acute{\rm e}$n speeds for different threads in a specific AR prominence studied by \cite{Okamoto2007Sci}, ranges between 120 to 150 km s$^{-1}$. Therefore, prominence seismology is a powerful tool for us to obtain the physical parameters in the promineces.

To comprehend the small-scale dynamics in solar prominences and the nature of vertical threads, we present noticeable transverse swaying motions of vertical threads in a solar prominence at the onset of its eruption. Combining data from the New Vacuum Solar Telescope (NVST), Chinese H$\alpha$ Solar Explorer (CHASE), and Solar Dynamics Observatory (SDO), we conduct a comprehensive study on the dynamic characteristics of transverse swaying motion in a solar prominence. Additionally, we employ prominence seismology to investigate the magnetic structure of the vertical threads. The sections of this letter are organized as follows: the observations and data reduction are described in Section \ref{sec:observations and methods}, the results are given in Section \ref{sec:results}, and a summary and discussions are presented in Section \ref{sec:summary and discussions}.

\section{Observations and data reduction} \label{sec:observations and methods}
The interested solar prominence was located on the northeast solar limb at a solar latitude of around 80$\degr$ (-200$\arcsec$, 920$\arcsec$) on May 11, 2023 (see Fig.\ref{fig1} (a)). In order to better display the prominence, we rotate the images clockwise by 14.5$\degr$ to align the solar limb horizontally (see Fig.\ref{fig1} (b)-(d)). This prominence was observed by the New Vacuum Solar Telescope\footnote{\url{http://fso.ynao.ac.cn}} (NVST; \citealp{Liu2014RAA,Yan2020ScChE}), Chinese H$\alpha$ Solar Explorer \footnote{\url{https://ssdc.nju.edu.cn}}(CHASE; \citealp{Li2019RAA,Lichuan2022SCPMA}), and Solar Dynamics Observatory\footnote{\url{https://sdo.gsfc.nasa.gov}} (SDO; \citealp{Pesnell2012SoPh}). NVST is a vacuum solar telescope with a 985 mm clear aperture located at Fuxian Lake, in Yunnan Province, China, which mainly includes a imaging system, an adaptive optics (AO) system \citep{Zhang2023SCPMA} and a spectrometer for now. The imaging system mainly includes H$\alpha$ and TiO-band channels. The H$\alpha$ channel is equipped in a tunable Lyot filter with a band width of 0.25 $\rm\AA$, which can record the images in the $\pm$ 5 $\rm\AA$ range with a step size of 0.1 $\rm\AA$.
The high-resolution images of the prominence at the H$\alpha$ center recorded by the imaging system are utilized in our study. The field of view of these H$\alpha$ images is 150$\arcsec$ $\times$ 150$\arcsec$, with around 12 s cadence and a CCD plate scale of 0.165$\arcsec$ per pixel. All NVST H$\alpha$ images are normalized by the quiet Sun, marked by yellow box in panel (b) of Fig.\ref{fig1}, to  diminish the effect of seeing, and aligned with each other based on a cross-correlation algorithm \citep{Yang2015RAA}. 
The H$\alpha$ Imaging Spectrograph (HIS) on board the CHASE satellite can provide solar spectra in wavebands of H$\alpha$ (6559.7-6565.9 $\rm\AA$) and Fe I (6567.8-6570.6 $\rm\AA$) with high spectral and temporal resolutions. The CCD plate scale of HIS along the slit is 0.52$\arcsec$ per pixel, and the scanning step is also 0.52$\arcsec$. The length of the slit is 23 mm, and the width is 9 $\mu$m. The full-disk scanning time is 60 seconds, and the spectral resolution is 0.024 $\rm\AA$ per pixel. Full-disk scanning H$\alpha$ spectrum from the H$\alpha$ Imaging Spectrograph (HIS) on board the CHASE satellite is utilized to diagnose the physical parameter in the prominence. The spectrum data from CHASE/HIS are calibrated from Level 0 to Level 1, including the dark-field and flat-field correction, slit image curvature correction, wavelength and intensity calibration, and coordinate transformation \citep{Qiu2022SCPMA}.`
For better showing, we reverse the intensity of H$\alpha$ observations of the NVST and CHASE (see panels (b) \& (d) of Fig.\ref{fig1}). The extreme-ultraviolet (EUV) 171 $\rm \AA$ and 193 $\rm\AA$ images obtained from Atmospheric Imaging Assembly (AIA; \citealp{Lemen2012SoPh}) on board SDO are also used to show the prominence in our study. The CCD plate scale and cadence are 0.6$\arcsec$ per pixel and 12 seconds, respectively.
Results \label{sec:results}
\subsection{Swaying motions and upward flow observed in the prominence} 
Fig.\ref{fig1} shows the observations of the prominence at around 02:52 UT. The prominence is noticeable in absorbing structures observed at EUV wavelengths of EUV 193 $\rm\AA$ and 171 $\rm\AA$ (see panels (a)\&(c)), while appearing bright in H$\alpha$ observations (see panels (b)\&(d)). Notably, the H$\alpha$ observations is reversed, indicated by a negative sign. Numerous vertical threads within the prominence are discernible. These vertical threads were rooted on the solar disk nearby the solar limb. It should be noted that a bright feature appeared near the base of these vertical threads, marked by the blue arrow in panel (c) and circles in the animation of Fig.\ref{fig1}, suggesting a heating event in that region. An eruption took place on the right-hand side close to the prominence around 03:00 UT, which may be related to the subsequent eruption of the prominence. Following this event, the prominence became active and unstable, and then began to erupt, as evident in the animation of Fig.\ref{fig1}. During the onset of prominence eruption, some vertical threads exhibited a transverse swaying motion, accompanied by the lifting of material along the vertical threads from their base. Figs.\ref{fig2} (b)\&(c) exhibit the swaying motion of the vertical thread at two different moments, with the black dotted lines outlining the thread's boundary. The conspicuous transverse wave-like structures are clearly identifiable. The half of the projected wavelength of the wave-like structure in panel (b) is determined to be 11.6 Mm while the projected wavelength of the wave-like structure in panel (c) is derived to be 22.5 Mm. Therefore, the mean projected full wavelength of these two wave-like structures is approximately 22.9 Mm. 
These transverse motions can be considered signatures of transverse MHD kink waves travelling along the prominence threads \citep{Lin2009ApJ, Li2022SCPMA}. Simultaneously, we also observe some materials flowing along these swaying threads from their base, which is associated with the heating event. Panel (d) displays the time-distance diagram reconstructed by a series of NVST H$\alpha$ images along the blue line AB in panel (a). In this diagram, inclined stripes representing the flowing material are marked by the inclined dotted lines, and their velocities are determined to be approximately 26.7 km s$^{-1}$, 26.9 km s$^{-1}$, 28.9 km s$^{-1}$, and 26.2 km s$^{-1}$, with a mean velocity of about 27 km s$^{-1}$. This also implies that the vertical threads of the prominence were experiencing significant velocity shear. Additionally, it is found that the speed of flowing material in certain locations exhibits an increasing trend, as indicated by the blue dotted lines in panel (d).

\subsection{Characteristics of the swaying motions}
To conduct a detailed study of these swaying motions representing MHD kink waves, we make time-distance diagrams along the path perpendicular to their propagation direction. Figs.\ref{fig3} (a)\&(b) display the time-distance diagrams along the line CD in Fig.\ref{fig2} (a), reconstructed by NVST H$\alpha$ images and SDO/AIA 193 $\rm\AA$ images, respectively.
From panels (a)\&(b) and the animation in Fig.\ref{fig3}, numerous periodic oscillations in bright or dark structures can be identified, representing the signals of MHD kink waves. Many oscillations can be identified over several periods, and they have a comparable oscillation period (see the animation in Fig.\ref{fig3}). It should be noted that the bright signals in the H$\alpha$ and 193 $\rm\AA$ channels are considered as the gap between two vertical threads.
Panels (c1)-(c3) and (d1)-(d3) display several time-distance diagrams along the three yellow solid lines in Fig.\ref{fig1} (a), representing different heights of the vertical threads. Panels (c1)-(c3) are reconstructed from NVST H$\alpha$ observations, while panels (d1)-(d3) correspond to SDO/AIA 193 $\rm\AA$. 
Based on panels (c1) and (c2), the width of the dark stripes can be identified as approximately 0.5 Mm.
There are many remarkable and similar oscillating signals in both time-distance diagrams.
The dark and bright structures are coupled in these oscillating signals, and the bright signals are more distinct in H$\alpha$ images. Given this fact, we consider these bright signals as representing the swaying motions or MHD kink waves of the vertical threads.
The yellow dots in panel (c2) outline one signal of the transverse swaying motion with an entire period, which has been plotted in the corresponding SDO/AIA 193 $\rm\AA$ image in panel (d2). The signal of the swaying motions in 193 $\rm\AA$ observations is thicker than that in H$\alpha$ observations, possibly due to the lower spatial resolution in 193 $\rm\AA$ observations. By identifying the peaks and troughs of these signals (marked by different white vertical solid lines), we could estimate their periods to be 16.2, 13.8, and 14.4 minutes, respectively. These periods are comparable to the ones (10-16 minutes) of oscillations found in a quiescent prominence by \cite{Li2022SCPMA}.
% but longer than the ones (130-250 seconds) found by \cite{Okamoto2007Sci,Okamoto2015ApJ} and the ones (2-5 minutes) found by \cite{Lin2007SoPh}. This discrepancy may be associated with different magnetic or plasma characteristics in different prominences.

On the other hand, we make a in-depth analysis of the swaying motions associated with panels (c1) and (c2). 
Fig.\ref{fig4} illustrates the evolution of the swaying motion associated with panel (c1), named as SMa, while Fig.\ref{fig41} is for the evolution of the swaying motion associated with panel (c2), named as SMb. Panels (a1)-(a6) of Figs.\ref{fig4}\&\ref{fig41} display time-distance diagrams along six parallel cuts at a distance of 0.55 Mm from each other, as indicated by six light blue and red lines in Fig.\ref{fig2} (a), representing heights from higher to lower positions. This setup could capture the propagation of a single transverse swaying motion passing through different heights from bottom to top along the vertical threads. 
It is found that there was a time delay of the peaks among these wave structures marked by the yellow and blue plus sign in Figs.\ref{fig4}\&\ref{fig41}, suggesting that these swaying motions correspond to traveling waves, a phenomenon also observed in coronal loops \citep[e.g.,][]{Li2023A&A}. According to the peaks of these wave structures, we plot the height of the peaks for each swaying motion, represented by the height of each cut with respect to time, as shown in the white box in panels (a6) in Figs.\ref{fig4}\&\ref{fig41}. Subsequently, we fit these data points with a linear line outlined by the yellow lines. The inclination of the fitted line represents the phase speed of this traveling wave, and we obtain that the phase speeds of SMa and SMb are 23.0$\pm$2.4 km s$^{-1}$ and 29.0$\pm$3.9 km s$^{-1}$, respectively. These phase velocities are comparable to the projected velocities of the flowing materials. Panels (b1)-(b6) of Figs.\ref{fig4}\&\ref{fig41} outline the displacements of corresponding wave-like structures. We fit them by using the following equation:
\begin{equation}
   D(t)=A_m sin(\frac{2\pi}{P}t+\varphi)+kt+D_0,
\end{equation}

\noindent in which $A_m$, $P$, and $\varphi$ represent the displacement amplitude, the oscillatory period, the initial phase of the oscillatory motion, respectively. $k$ and $D_0$ are fitting parameters associated with a linear background. Table 1 exhibits the key parameters of the transverse swaying motion. The uncertainties arise from the estimated displacements, which are regarded as two pixels in the images.

\begin{table}[!h]
    \centering
    \caption{Key parameters of the transverse swaying motions of SMa\footnote{ $D_{a1}-D_{a6}$ correspond to panels (b1)-(b6) of Fig.\ref{fig4} for SMa. } and SMb\footnote{ $D_{b1}-D_{b6}$ correspond to panels (b1)-(b6) of Fig.\ref{fig41} for SMb.}}
    \begin{tabular}{|c|ccccc|c|c|}
       \hline
       parameter  & $A_m$ (Mm) & $P$ (min) & $\varphi$ & $k$ & $D_0$ & velocity amplitude (km s$^{-1}$) & wavelength (Mm) \\
       \hline
       $D_{a1}$  & $0.58\pm0.04$  & $17.5\pm0.3$ & $2.7\pm1.7$ & $-2.0\pm0.2$ & $10.0\pm0.7$ & $3.47\pm0.30$ & $24.2\pm2.9$ \\
       \hline
       $D_{a2}$  & $0.54\pm0.05$ & $16.4\pm0.4$ & $3.6\pm2.3$ & $-2.1\pm0.2$  & $10.0\pm1.0$ & $3.45\pm0.40$ & $22.6\pm2.9$\\
       \hline
       $D_{a3}$  & $0.53\pm0.05$  & $16.3\pm0.4$ & $3.3\pm2.1$ & $-2.1\pm0.2$   & $10.2\pm0.9$ & $3.41\pm0.40$ & $22.5\pm2.9$\\
       \hline
       $D_{a4}$ & $0.52\pm0.05$  & $15.2\pm0.5$  & $3.5\pm2.9$ & $-1.8\pm0.3$  & $9.1\pm1.2$ & $3.58\pm0.46$ & $21.0\pm2.9$\\
       \hline
       $D_{a5}$ & $0.45\pm0.04$ & $15.4\pm0.4$ & $5.6\pm2.7$ & $-1.4\pm0.2$ & $7.7\pm0.9$ & $3.06\pm0.36$ & $21.3\pm2.8$\\
       \hline
       $D_{a6}$ & $0.39\pm0.04$ & $14.8\pm0.5$ & $1.5\pm3.3$ & $-0.7\pm0.2$ & $5.2\pm0.9$ & $2.76\pm0.38$ & $20.4\pm2.8$\\
       \hline\hline
       $D_{b1}$ & $0.59\pm0.05$ & $14.5\pm0.4$  & $3.0\pm2.7$ & $-2.1\pm0.3$ & $10.5\pm1.2$ & $4.26\pm0.48$ & $25.2\pm3.9$ \\
       \hline
       $D_{b2}$ & $0.56\pm0.05$ & $14.5\pm0.4$  & $3.4\pm3.0$ & $-2.4\pm0.3$ & $11.5\pm1.2$ & $4.04\pm0.47$ & $25.2\pm4.1$\\
       \hline
       $D_{b3}$ & $0.54\pm0.05$ & $14.1\pm0.4$  & $0.5\pm3.2$ & $-1.9\pm0.3$ & $9.9\pm1.2$ & $4.01\pm0.49$ & $24.5\pm4.0$\\
       \hline
       $D_{b4}$ & $0.51\pm0.05$ & $14.4\pm0.5$  & $3.0\pm3.4$ & $-2.1\pm0.3$ & $10.8\pm1.2$ & $3.71\pm0.49$ & $25.1\pm4.2$\\
       \hline
        $D_{b5}$ & $0.48\pm0.05$ & $13.9\pm0.5$  & $5.2\pm3.4$ & $-2.2\pm0.3$ & $11.3\pm1.1$ & $3.62\pm0.51$ & $24.2\pm3.9$\\
        \hline
        $D_{b6}$ & $0.42\pm0.05$ & $13.8\pm0.5$  & $4.4\pm3.9$ & $-1.9\pm0.3$ & $9.8\pm1.1$ & $3.19\pm0.49$ & $24.0\pm4.1$\\
        \hline 
    \end{tabular}
    \label{tab1}
\end{table}

According to Table 1, the projected displacement amplitudes ($A_m$) and oscillatory periods ($P$) of SMa range from 0.39$\pm$0.04 to 0.58$\pm$0.04 Mm and from 14.8$\pm$0.5 to 17.5$\pm$0.3 minutes, respectively. For SMb, these values range from 0.42$\pm$0.05 to 0.59$\pm$0.05 Mm and from 13.8$\pm$0.5 to 14.5$\pm$0.4 minutes, respectively. Therefore, the projected velocity amplitudes ($v_{am}$=2$\pi *A_m/P$) can be derived (see the seventh column of Table 1), ranging from 2.76$\pm$0.38 to 3.47$\pm$0.30 km s$^{-1}$ with a mean of 3.29$\pm$0.38 km s$^{-1}$ for SMa and from 3.19$\pm$0.49 to 4.26$\pm$0.38 km s$^{-1}$ with a mean of 3.80$\pm$0.49 km s$^{-1}$ for SMb. These velocity amplitudes are consistent with the findings of previous studies \citep{Lin2009ApJ, Li2022SCPMA}, confirming that these transverse swaying motions are classified as small-amplitude oscillations. On the other hand, the ratio of the displacement amplitude to the thread's cross-section radius (approximately 0.25 Mm, as seen in Fig.\ref{fig1} (b) and Fig.\ref{fig3} (c1)\&(c3)) is larger than unity. Therefore, these transverse waves exhibit non-linear oscillations \citep{Ruderman2010PhPl,Ruderman2014SoPh}. An intriguing finding is that $A_m$, $P$, and $v_{am}$ exhibit an increasing trend from $D_{a6}$/$D_{b6}$ to $D_{a1}$/$D_{b1}$.
It is worth noting that the positions of $D_{a6}$/$D_{b6}$ to $D_{a1}$/$D_{b1}$ represent a progression from lower to higher cuts as the swaying motions propagate upward. This implies that the projected displacement amplitudes, oscillatory periods, and projected velocity amplitudes of SMa and SMb increase over time.
Combining the previous finding that there appears to be a significant increase in speed in some locations (marked by blue dotted lines in panel (d) of Fig.\ref{fig2}), we speculate that the increase in flow speed could lead to an increase in the amplitude of the swaying motion. This phenomenon would be linked to shear flow and negative-energy wave instabilities \citep{Ryutova1988JETP,Ruderman1995JPlPh}.  
 
Based on the oscillatory period (P) and projected phase velocity($V_{pph}$), it also could be derived the projected wavelengths ($P*V_{pph}$), which range from 20.4$\pm$2.8 to 24.2$\pm$2.9 Mm for SMa and from 24.0$\pm$4.1 to 25.2$\pm$3.9 Mm for SMb. It also exhibits an increasing pattern over time. The mean projected wavelength is 23.3$\pm$3.5 Mm, which is consistent with that (22.9 Mm) derived by the above direct measure. It also confirms that the bright signals could be used to represent the swaying motions ro MHD kink waves of the vertical threads.
It is also found that SMb exhibits a larger projected velocity amplitude and phase velocity in comparison to SMa. This suggests that higher transverse swaying motion in the vertical threads corresponds to a larger projected velocity amplitude and phase velocity. Such variations indicate the presence of distinct local magnetic or plasma properties at different heights within the vertical threads and also imply that the prominence has been subject to significant velocity shear.

\subsection{The line-of-sight (LOS) velocities in the prominence}

Fig.\ref{fig5} exhibits the observations from CHASE/HIS at around 04:07 UT. Panel (a) displays the composite map constructed using a set of scanning spectrum data at H$\alpha$ center from 04:07:29 to 04:07:33 UT. This moment is indicated by the white vertical line in Fig.\ref{fig2} (d), coinciding with the upflowing material. Panel (b) shows the H$\alpha$ spectrum along the slit marked by the black line on panel (a), which is located in the prominence. To better display the prominence spectrum, we also reverse the intensity in the spectrum map. 
Panel (c) displays the corresponding map of full width at half maximum (FWHM) of the profiles, taken at the same observational time as panel (a). We utilize the mean intensity within the wavelength range from 6560.4 $\rm\AA$ to 6561.4 $\rm\AA$, and from 6564.2 $\rm\AA$ to 6565.0 $\rm\AA$ as the baseline for the line profiles. The FWHM is calculated as the width of the profile where the intensity exceeds half of the maximum amplitude. The FWHM in the prominence region is almost less than 1 $\rm\AA$. The blue solid line corresponds to the line AB in Fig.\ref{fig2} (a), which is for the time-distance diagram of Fig.\ref{fig2} (d). Two yellow asterisks on the line correspond to two yellow asterisks in Fig.\ref{fig2} (d). Therefore, we consider that the plasma in the white box could be related to flowing material. The FWHM in the white box is determined to be 0.72$\pm$0.03 $\rm\AA$, while in the higher region marked by the red box, where lower optical thickness is expected, it is 0.61$\pm$0.03 $\rm \AA$. The errors are represented by standard deviations. The similar values of the FWHMs in these two regions suggest that the profile of the white box could be regarded as under the optically thin approximation hypothesis.
Panel (d) shows the average intensity of the spectrum over the white box, with error bars are represented by corresponding standard deviations. 
The shape of the profile (diamond) is mostly single-peaked, which supports the optically thin assumption.
The asymmetry of the profile also reveals a background containing some additional red-shifted components. 
Based on the optically thin approximation hypothesis, we use a single and double Gaussian functions to fit the spectrum curve, respectively. 
The red dashed line represents the result derived by a single Gaussian function, with a Doppler velocity of 4.07$\pm$0.07 km s$^{-1}$. The blue lines represent the result derived by a double Gaussian function, where the two dotted blue lines correspond to the slow and fast components, and the solid blue line corresponds to the combination of the two components. It can be derived that the Doppler velocities of the slow and fast components are -0.21$\pm$1.83 km s$^{-1}$ and 16.37$\pm$2.44 km s$^{-1}$.
We believe that these fast components are closely associated with the upflowing plasma. Therefore, the line-of-sight velocities of the fast components are considered to represent the line-of-sight velocities of the upflowing plasma.

Next, by combining the projected velocity on the sky plane of approximately 27 km s$^{-1}$, we can deduce that the full velocity of the upflow plasma is 31.5 km s$^{-1}$, with an inclination angle of 31$\degr$ relative to the plane of the sky. Considering that these lifting materials propagated along the vertical threads, it can be inferred that the vertical threads have an inclination angle of 31$\degr$ relative to the plane of the sky. On the other hand, the transverse swaying motions also propagated along these vertical threads. Then, the velocity amplitude and phase velocity can be corrected for the projected effect. For SMa, the corrected mean velocity amplitude and phase velocity are 3.84$\pm$0.44 km s$^{-1}$ and 26.9$\pm$2.8 km s$^{-1}$. For SMb, the corrected velocity amplitude and phase velocity are 4.44$\pm$0.57 km s$^{-1}$ and 33.9$\pm$4.6 km s$^{-1}$. The mean wavelength of these oscillatory motions can be corrected to be about 28.4$\pm$4.3 Mm.

\subsection{Seismology on the transverse MHD kink mode waves}

As mentioned earlier, these transverse swaying motions can be considered as signatures of transverse MHD kink mode waves \citep{Okamoto2007Sci, Lin2009ApJ, Li2022SCPMA}. Assuming that these prominence threads are composed of numerous magnetic flux tubes configured as straight and homogeneous cylinders filled with filament-like plasma, and given that the density within the prominence is significantly higher than the external coronal density, the Alfv$\acute{\rm e}$n speed ($v_A$) in the thread can be estimated using the following equation \citep{Lin2009ApJ}:
\begin{equation}
    v_A=V_{ph}/\sqrt 2,
\end{equation}
\noindent where $V_{ph}$ is the phase velocity of the MHD kink mode wave represented by transverse swaying motions. Using the corrected phase velocities, we can estimate the Alfv$\acute{\rm e}$n speeds to be around 19.0$\pm$2.0 km s$^{-1}$ and 24.0$\pm$3.3 km s$^{-1}$ for SMa and SMb, respectively. It should be noted that SMa and SMb are located in the same series of vertical threads at different heights, with SMa below SMb. The difference in Alfv$\acute{\rm e}$n speed between the two MHD kink mode waves may be related to the varying plasma density at different heights within the same vertical threads. 
Assuming that the same vertical threads (the same magnetic flux tubes) have similar magnetic field strength and that the magnetic field strength does not vary much between the two heights, we can utilize the Alfv$\acute{\rm e}$n speed equation ($v_A = B / \sqrt{ \mu_0 \rho}$) to deduce that the density in SMa is approximately 1.6 times denser than that in SMb.

On the other hand, assuming a density similar to that of prominences or coronal rains for the thread, within the range of $\rho = (1.67-16.7) \times 10^{-11}$ kg m$^{-3}$ \citep{Stellmacher1997A&A,Labrosse2010SSRv,Wang2018ApJ_863_180,Antolin2021NatAs}, and considering a mean Alfv$\acute{\rm e}$n velocity of 21.5 km s$^{-1}$ within the threads, the magnetic field strength of these threads is estimated to be approximately 1-3 Gauss. The time-averaged wave energy flux ($<F>$) carried by these kink MHD waves in multiple magentic flux tube structures can be estimated by the following equation \citep{Morton2012NatCo,Goossens2013ApJ,Doorsselaere2014ApJ},

\begin{eqnarray}
  <F> & =&\frac{1}{4}*f*V_{ph}(\rho v_{am}^2+\frac{b^2}{\mu_0}) \\ 
  b &\approx& B \frac{2\pi}{\lambda}A_m 
\end{eqnarray}
\noindent where $f$, $V_{ph}$, $v_{am}$, $b$, $\mu_0$, $\lambda$, and $A_m$ represent filling factor, the phase velocity, velocity amplitude, Lagrangian perturbation of the magnetic field, magnetic permeability of vacuum, wavelength of oscillation, and displacement amplitude, respectively. \cite{Doorsselaere2014ApJ} proposed that the energy flux in multiple kink waves should be corrected with a filling factor. Taking the parameter values derived from the above analysis ($V_{ph}$ = 30.4 km s$^{-1}$, $\rho = (1.67-16.7) \times 10^{-11}$ kg m$^{-3}$, $v_{am}$ = 4.14 km s$^{-1}$, $B$ = 1-3 G, $\lambda$ = 28.4 Mm, $A_m$ = 0.59 Mm, $\mu_0$ = 4$\pi \times 10^{-7}$ N/A$^2$) and considering a unit filling factor for these prominence threads, the energy flux carried by these kink MHD waves could be estimated to be approximately (0.32-3.1) $\times$ 10$^4$ erg s$^{-1}$ cm$^{-2}$. On the other hand, when also considering the correction factor for filling, the kinetic energy flux of the upward flow could be calculated by the equation of $F_{fl}$ = $\frac{1}{2}\rho v_{fl}^2*v_{fl}*f$, where $v_{fl}$ is the corrected velocity (31.5 km s$^{-1}$) of upflow material. Then the energy flux of the upward flow could be estimated to be approximately (5.2-52.2) $\times$ 10$^4$ erg s$^{-1}$ cm$^{-2}$. We can see that the energy carried by these swaying motions is only a few percent of that of the upward flow. 

\section{Summary and discussions} \label{sec:summary and discussions}

In this letter, we analyse a novel phenomenon where swaying motions occur in the vertical threads of the prominence during the onset of its eruption on May 11, 2023, utilizing high-resolution images and spectrum observations from NVST, CHASE, and SDO. The main results are presented as follows.

1. Transverse swaying motions in the vertical threads of the prominence travel upward from the base of the threads, accompanied by upflowing materials with a mean velocity of 31.5 km s$^{-1}$ and an inclination angle of 31 degrees relative to the plane of the sky.

2. Transverse swaying motions exhibit oscillation patterns with a corrected velocity amplitude of around 3-4 km s$^{-1}$. The amplitudes and periods are estimated to be about 0.4-0.6 Mm and 13-17 minutes, respectively, while their phase velocities are estimated to be about 26-34 km s$^{-1}$.
Furthermore, the oscillations of swaying motion exhibit an increasing pattern over time in displacement amplitudes, oscillatory periods, and projected velocity amplitudes, which are considered to be related to the increasing upward flow in some locations.

3. Kink MHD waves are proposed as an explanation for these transverse swaying motion in the threads. Utilizing prominence seismology and measured values of density, the Alfv$\acute{\rm e}$n speed and magnetic field strength in the vertical threads have been estimated to be around 21.5 km s$^{-1}$ and 1-3 Gauss, respectively. The energy carried by these swaying motions has been estimated to be (0.32-3.1) $\times$ 10$^4$ erg s$^{-1}$ cm$^{-2}$.

Most previous studies on this type of oscillatory motion have focused on horizontal threads of quiescent prominences/filaments \citep{Okamoto2007Sci, Lin2009ApJ}. Our study unveils similar oscillatory behaviors in the vertical threads of prominences. This observation may further demonstrate that the vertical threads in limb prominences are composed of aligned and interconnected magnetic structures, rather than separate magnetic dips of horizontal field \citep{Chae2010ApJ}. In the case of separate magnetic dips with a horizontal field, one would expect kink waves to propagate horizontally along the magnetic field lines. This propagation should induce vertically polarised transverse oscillations within the threads, rather than horizontally polarised transverse transverse oscillations, which contradicts what we have observed. On the other hand, some upflows propagating smoothly along these vertical threads also provide evidence for the vertical field structure of these threads.
Additionally, in combination with the Doppler velocity derived from CHASE observations, we determine that the upflowing materials have an inclination angle of 31 degrees relative to the plane of the sky. These upflowing materials are accompanied by transverse swaying motion in the prominence threads. These features collectively support the notion that the vertical threads viewed in limb prominences have an inclination angle of 31 degrees relative to the plane of the sky. 
Based on the prominence seismology, the magnetic field strength of the prominence threads has been estimated to be around 1-3 Gauss in our study, which is consistent with previous discoveries in fine thread-like structures observed in solar quiescent filaments/prominences \citep{Lin2009ApJ,Schmieder2013ApJ,Li2022SCPMA}.

In terms of the origin of the oscillations, the observation of these swaying motions accompanied by upward flows suggests that these oscillations may be induced by shear flows associated with negative-energy waves \citep{Ryutova1988JETP,Ruderman1995JPlPh}. Although the flow speed (30-34 km s$^{-1}$) slightly exceeds the obtained maximum Alfv$\acute{\rm e}$n speed (24.0$\pm$3.3 km s$^{-1}$) within the threads, considering a thin external circumstances, it would require a shear velocity significantly greater than 34 km s$^{-1}$ to trigger the Kelvin-Helmholtz instability in our case \citep{Zaqarashvili2015ApJ_813_123}. Furthermore, during the transverse swaying motions, the vertical threads maintained their shapes without exhibiting any vortex motions (see animation 1), indicating the absence of Kelvin-Helmholtz instability. Therefore, we consider that these swaying motions are signatures of negative-energy waves, when the shear velocity is smaller than the threshold value for the onset of the Kelvin-Helmholtz instability \citep{Ryutova1988JETP,Ruderman1995JPlPh}.
Furthermore, we find that the flow speed is comparable to the phase speed of the waves, potentially leading to negative-energy wave instabilities \citep{Nakariakov2021SSRv}. In such cases, the wave absorbs energy from the flow, resulting in an increase in amplitude, consistent with our finding that the oscillation amplitude increased over time. Therefore, we propose that this intriguing phenomenon is linked to negative-energy wave instabilities. As negative-energy wave instabilities occur, they may promote the instability and disruption of the prominence.

\begin{figure}[ht!]
\plotone{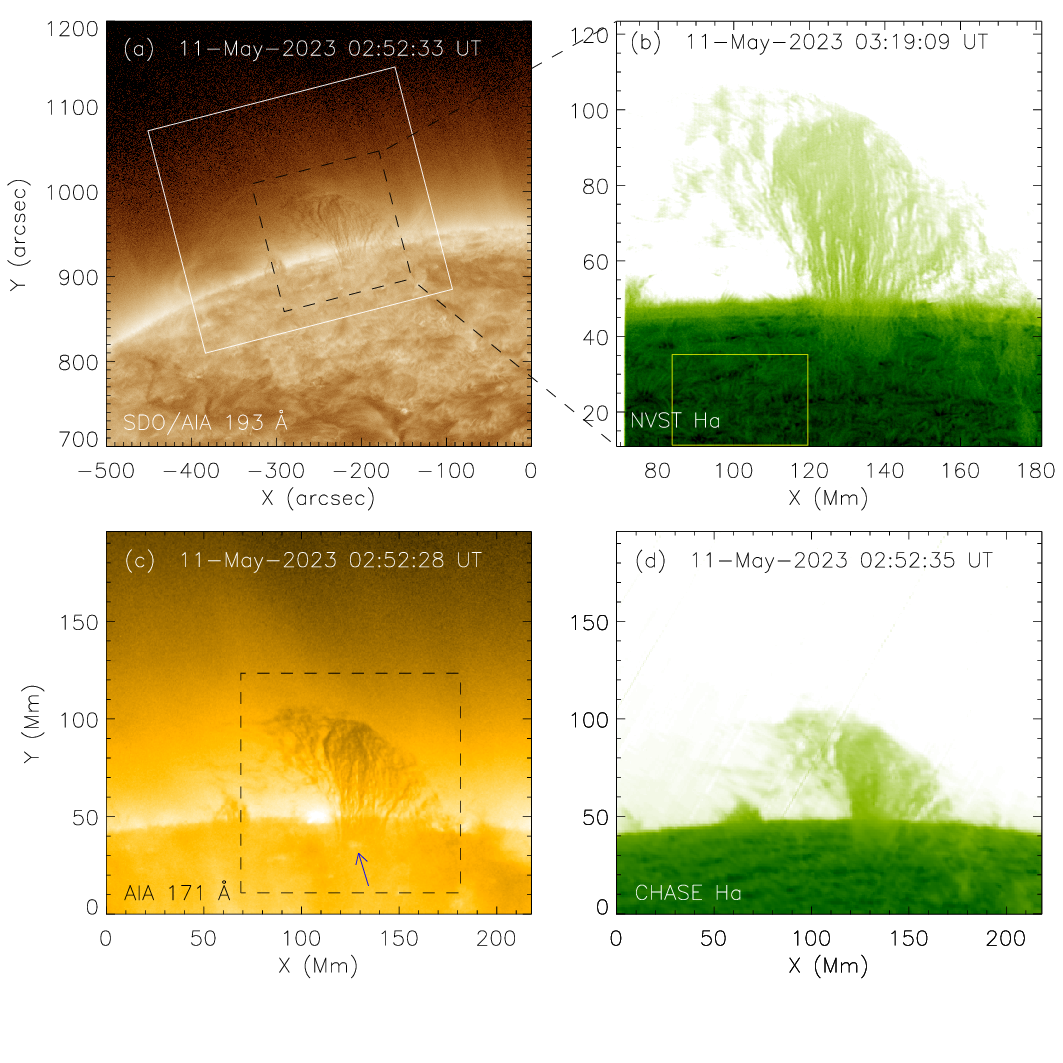}
\caption{Overview observations of the prominence. (a): SDO/AIA 193 $\rm\AA$ image. The black solid box outlines the field of view (FOV) of panels (c)\&(d). (b): H$\alpha$ images observed by NVST, with the FOV marked by dashed boxes on panels (a) and (c). (c): Corresponding SDO/AIA 171 $\rm\AA$ image. (d): H$\alpha$ composite image constructed by CHASE spectrum data. The FOV in panels (b)-(d) have been clockwise rotated by 14.5 degrees with respect to panel (a). To maintain a consistent coordinate system, the rotated images align with the coordinate system of panel (c). An animation is available, which shows the evolution of prominence from 01:11 UT to 03:45 UT on May 11, 2023.
\label{fig1}}
\end{figure}

\begin{figure}[t!]
\plotone{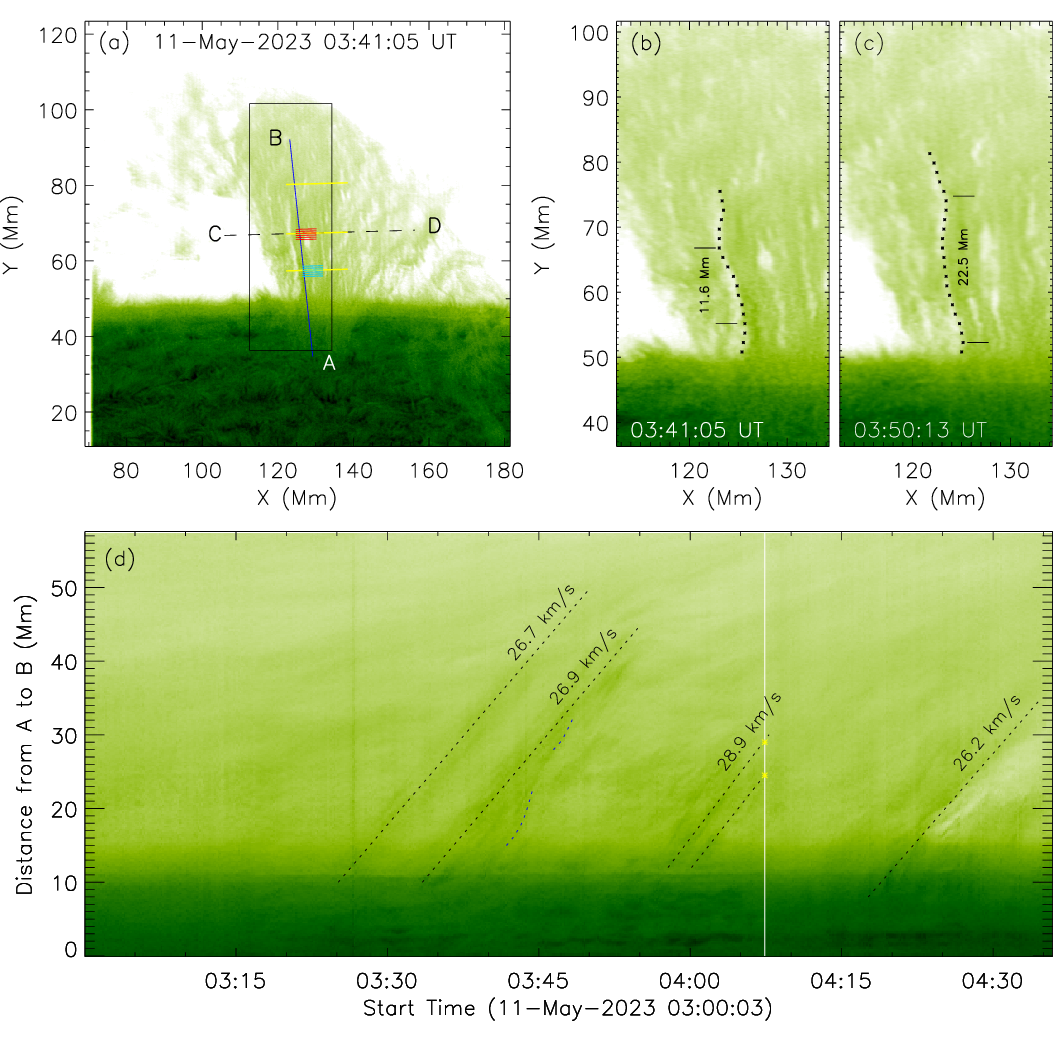}
\caption{Transverse swaying motion observed by NVST. (a): NVST H$\alpha$ image at 03:41:05 UT. The six light blue and six red lines indicate the parallel cuts for the time-distance diagram of Figs.\ref{fig4}\&\ref{fig41}. (b)-(d): Two zoom-in images of H$\alpha$ at different moments. Dotted lines mark the transverse swaying motion in the vertical threads. The FOV is outlined by the black box in panel (a). (d): Time-distance diagram constructed from a series of H$\alpha$ observations along the blue solid line AB in panel (a). The white vertical line denotes the moment of Fig.\ref{fig5}.
\label{fig2}}
\end{figure}

\begin{figure}[t!]
\plotone{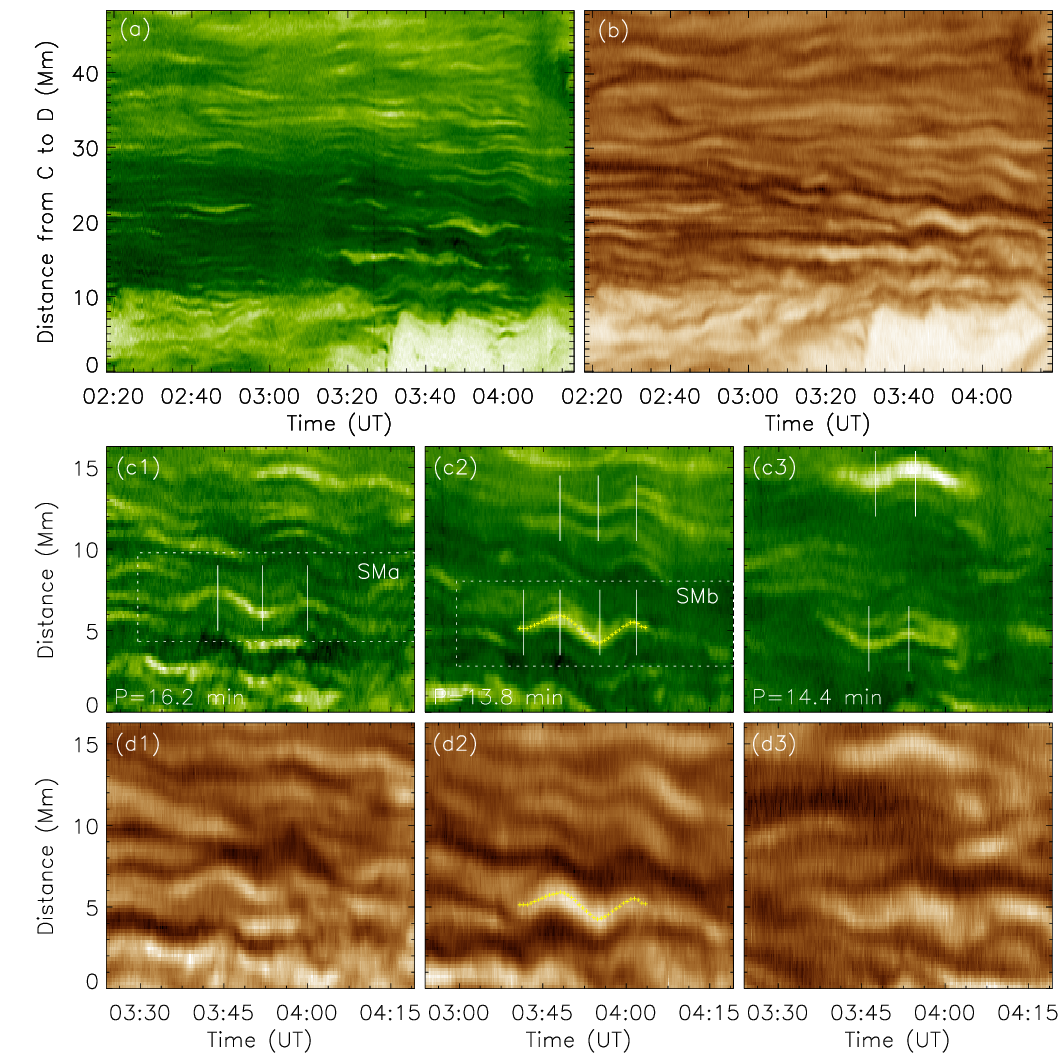}
\caption{Time-distance diagrams illustrating the transverse swaying motions. (a)-(b): Time-distance diagrams constructed from NVST H$\alpha$ and SDO/AIA 193 $\rm\AA$ images along the dashed line CD in Fig.\ref{fig2} (a). (c1)-(c3): Time-distance diagrams constructed from NVST H$\alpha$ images along three parallel yellow lines of Fig.\ref{fig2} (a) at different heights. The peaks and troughs of these oscillations are marked by different white vertical solid lines. (d1)-(d3): Corresponding time-distance diagrams constructed from SDO/AIA 193 $\rm\AA$ images. The yellow dots in panel (c2) outline one signal of the transverse swaying motion with an entire period, which has been plotted in the corresponding SDO/AIA 193 $\rm\AA$ image in panel (d2). An animation is available, which shows the time-distance diagrams along different height slit and also include a image of Fig.\ref{fig2}. The duration of the time-distance diagrams was from 02:18 UT to 04:18 UT.
\label{fig3}}
\end{figure}

\begin{figure}[ht!]
\plotone{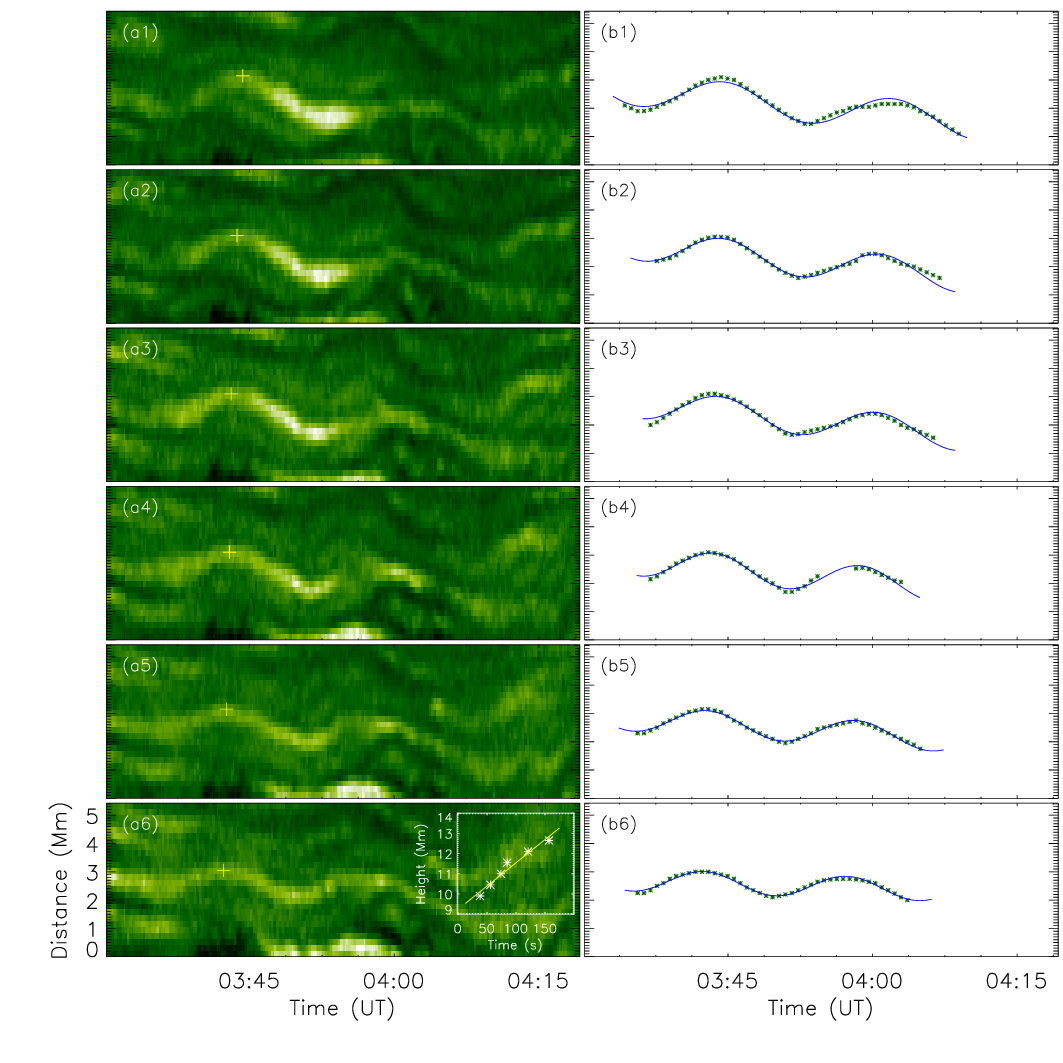}
\caption{Evolution of SMa. (a1)-(a6): Time-distance diagrams along different parallel cuts, spaced at a distance of 0.55 Mm from each other and from higher to lower positions. The field of view is outlined by the dotted box in Fig.\ref{fig3} (c1). Yellow plus signs mark the peaks of each transverse swaying motion. The white box in panel (a6) shows the profile of these peaks with respect to time, while the y-axis represents the height of each cut relative to each transverse swaying motion. The start time was at 03:41:39 UT. The yellow line denotes the fitted line with a linear function. (b1)-(b6): Profiles of each transverse swaying motion. The blue solid lines represent the fitted lines based on the function in Eq.(1).
\label{fig4}}
\end{figure}

\begin{figure}[ht!]
\plotone{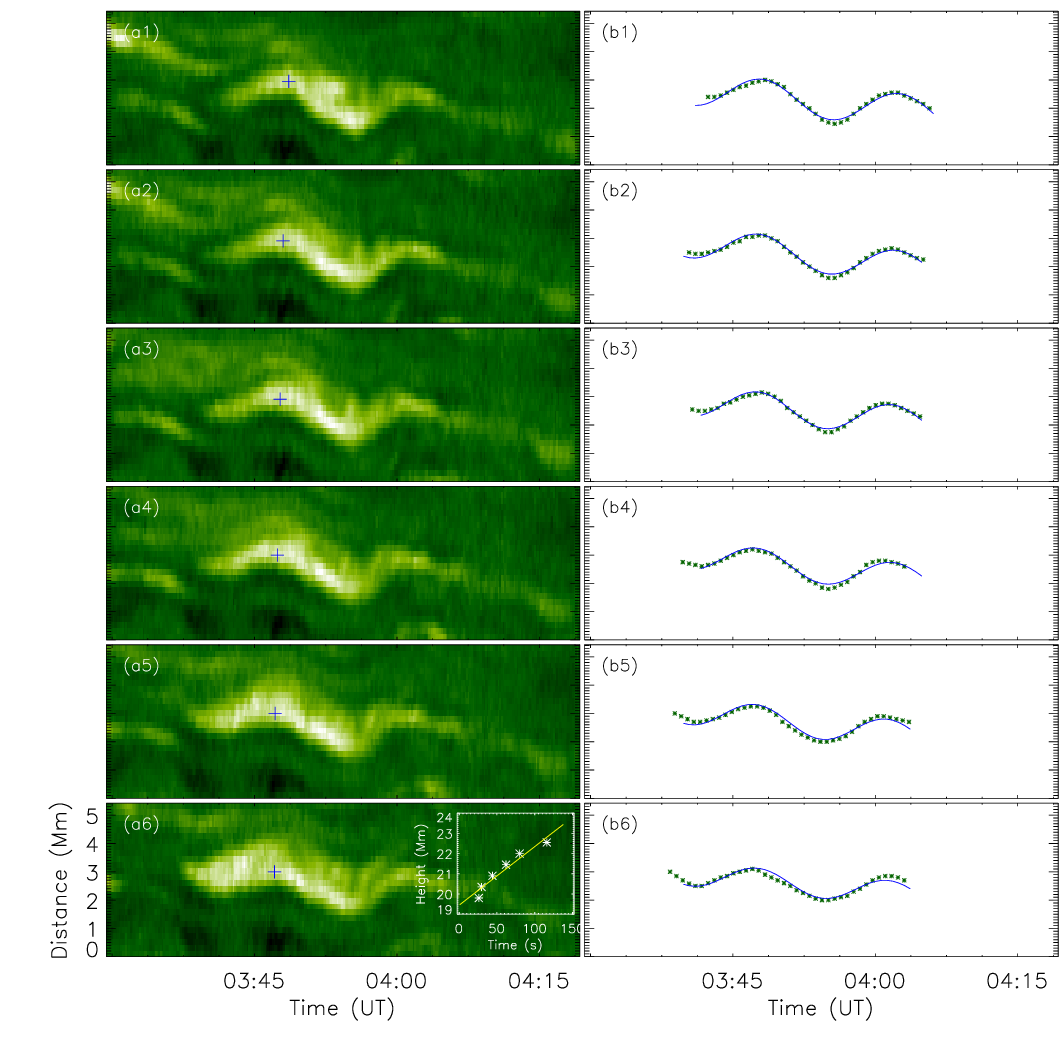}
\caption{Evolution of SMb. Same as Fig.\ref{fig4}, but for SMb. The field of view of panels (a1)-(ab1) is outlined by the dotted box in Fig.\ref{fig3} (c2). The start time of white box in panel (a6) was at 03:46:37 UT.
\label{fig41}}
\end{figure}

\begin{figure}[ht!]
\plotone{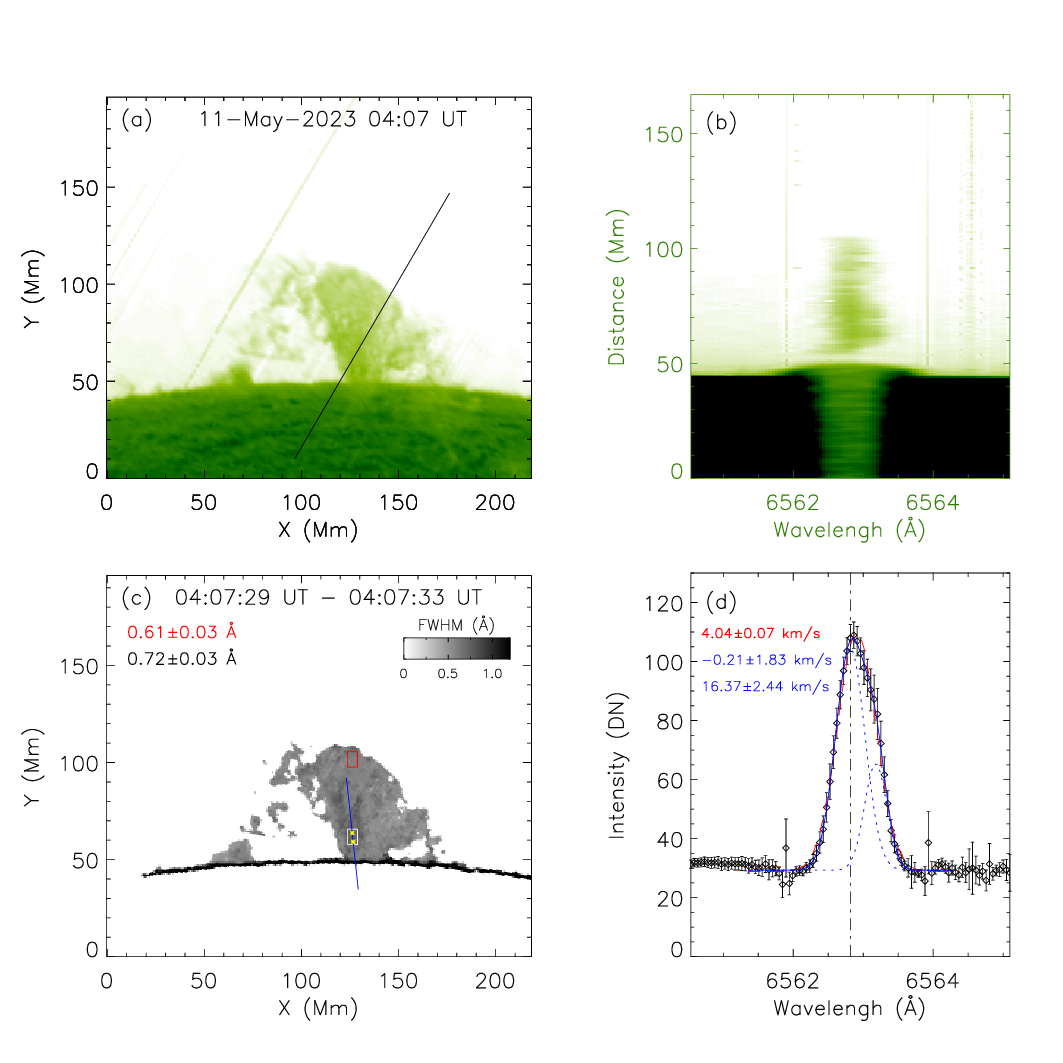}
\caption{Observations of CHASE/HIS H$\alpha$ spectrum data. (a): Composite map constructed using a set of scanning spectrum data at H$\alpha$ center at around 04:07 UT. (b): Spectrum image of H$\alpha$ along the black line on panel (a). (c): Corresponding FWHM map derived from the H$\alpha$ spectrum. The FWHM is determined to be the width of the profile where the intensity exceeds half of the maximum amplitude. The blue line corresponds to line AB in Fig.\ref{fig2} (a), while two yellow asterisks correspond to the two yellow asterisks in Fig.\ref{fig2} (d). (d): The profile of H$\alpha$ line at the region marked by the white box on panel (c). The H$\alpha$ line is calculated by the average over the white box while the error bars represent the corresponding standard deviation. The vertical dotted-dashed line denotes the H$\alpha$ center. The red dashed line and blue solid/dotted lines represent the fitted lines by single and double Gaussian functions, respectively. The red and blue values denote the Doppler velocity derived by single and double Gaussian function fitting, respectively.
\label{fig5}}
\end{figure}

What roles the energy carried by this type oscillation in the prominence play in the coronal heating, still remain controversial. In our study, the energy flux carried by the swaying motions is estimated to be approximately (0.32-3.1) $\times$ 10$^4$ erg s$^{-1}$ cm$^{-2}$. This value is comparable to the energy derived by \cite{Li2022SCPMA} and two orders of magnitude smaller than that derived by \cite{Okamoto2007Sci}, assuming a similar density. This difference may be attributed to the threads and waves studied by \cite{Okamoto2007Sci} being associated with an active region prominence, which is characterized by stronger magnetic fields. Otherwise, a higher energy flux of (1-2) $\times$ 10$^5$ erg s$^{-1}$ cm$^{-2}$ would be required to heat the quiet Sun corona \citep{Withbroe1977ARA&A,Aschwanden2007ApJ}. Therefore, the energy carried by the swaying motions may have only a minor impact on heating the quiet Sun corona. On the other hand, these swaying motions accompanied by upward flow appeared as the prominence began to lift during the onset of its eruption. Subsequently, the prominence collapsed rapidly following these swaying motions and upward flow (see Fig. \ref{fig2} (d) and the animation in Fig. \ref{fig1}). It is reasonable to believe that the energy carried by the upward flow and waves may have played a significant role in the disruption of the prominence.

\begin{acknowledgments}
We appreciate the referee's careful reading of the manuscript and many constructive comments, which helped greatly in improving the paper. The authors thank Prof. Ying Li at Purple Mountain Observatory and Prof. Jun Zhang at Anhui University for discussions and many valuable suggestions and comments. The authors are indebted to the NVST, CHASE, and SDO teams for providing the data. CHASE mission is supported by China National Space Administration (CNSA). 
SDO is a mission of NASA's Living With a Star Program. 
This work is supported by the Strategic Priority Research Program of the Chinese Academy of Sciences, Grant No.XDB0560000, the Youth Innovation Promotion Association CAS (2023063), 
the National Science Foundation of China (NSFC) under grant numbers 12003064, 12333009, 12003068, 12325303, 12203020, 12203097, 12273110, 12273060, the National Key R$\&$D Program of China (2019YFA0405000), the Yunnan Key Laboratory of Solar Physics and Space Science (202205AG070009), the Open Research Program of Yunnan Key Laboratory of Solar Physics and Space Science (YNSPCC202211), the Yunnan Science Foundation of China under number 202301AT070347, 202201AT070194, 202001AU070077, Yunnan Science Foundation for Distinguished Young Scholars No.202001AV070004, the grant associated with a project of the Group for Innovation of Yunnan province.
\end{acknowledgments}

\bibliography{references.bib}{}
\bibliographystyle{aasjournal}

\end{document}